\RequirePackage{lineno}
\documentclass[onecolumn,showpacs,preprintnumbers,amsmath,amssymb, prl,edmac]{revtex4}

\usepackage{dcolumn}
\usepackage{bm}
\usepackage{hyperref}
\usepackage{amsmath}
\usepackage{amsfonts}
\usepackage{graphics}
\usepackage{psfrag}
\usepackage{graphicx}
\usepackage{epsfig}
\usepackage{rotating}
 \usepackage[latin1]{inputenc}
 \usepackage{color}
 \usepackage{rotating}
 \usepackage{threeparttable}  
 \usepackage{multirow}        
 \usepackage{color}

\usepackage{lscape}

\usepackage{lineno}

\definecolor{DarkGreen}{rgb}{0,0.5,0}
\definecolor{Grey}{rgb}{0.5,0.5,0.5}
\definecolor{DarkYellow}{rgb}{1,0.7,0}
\definecolor{Violet}{rgb}{0.6,0.0,0.7}
\definecolor{Brown}{rgb}{0.5,0.3,0}

\begin{document}


\title{Observation of Evolutionary Velocity Field in Matching Pennies Game}
\author{Bin Xu }
\author{Zhijian Wang\footnote{Corresponding author, email: wangzj@zju.edu.cn}}

\affiliation{Experimental Social Science Laboratory, Zhejiang University, Hangzhou, 310058, China}


\date{\today}
\begin{abstract}
Matching pennies game is the simplest game in game theory. In data from laboratory experimental economics, with the metric for an instantaneous velocity measurement, we report the first observation of the evolutionary velocity field in the population strategy state space of the game.
\end{abstract}

\maketitle
\modulolinenumbers[1]
\linenumbers
\textbf{Background:} In evolutionary game theory (EGT), an evolutionary dynamic equation describes the velocity field. A velocity field quantifies the direction and the magnitude of the populations' strategy state change ($\triangle x$) in a unit of time ($\triangle t$) at each given state $x$ in the populations' strategy state space $\mathbb{X}$~\cite{Traulsen2009, Weibull1997,Bowles2004,Sigmund2010}.

Laboratory experimental economics removes EGT from its abstract
setting and links the theory to observed behavior~\cite{Samuelson2002}.
The observed populations behavior in the experiments are systematic, replicable, and roughly consistent with the dynamical systems approach~\cite{Crawford1991,Binmore1993, Huyck1995,Binmore2001, HuyckSamuelson2001, cason2005dynamics}.
The evolutionary velocity field has never before been obtained empirically.
In this letter, we will use a laboratory matching pennies game (MPG)  to experimentally obtain  the the
evolutionary velocity field.

\begin{figure}
\centering
\includegraphics[angle=0,width=6cm]{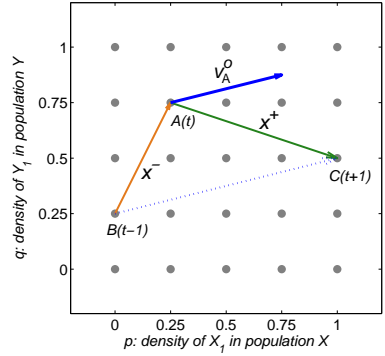}
\caption{Schematic diagram for an observation of an instantaneous velocity $v^{o}_{pq}$. The $5\times 5$ Lattice (in gray dots) is the population strategy state space $\mathbb{X}$ of the two populations MPG. $x_A$ is one of the states in $\mathbb{X}$ and $x_A$=$(0.25,0.75)$. Suppose that, during a microscope process $\mathbb{P}$
including both a forward change ($x^{+}$) and a backward chang ($x^{-}$),
the observed state at $t$ is $x_{A}$, and at ($t-1$) is $x_{B}$ and at ($t+1$) is $x_{C}$,
by Eq.~(\ref{eq:OneObsV}), $v^{o}_{A}$=$(x_C - x_B)/2$ (arrow in blue) is an instantaneous velocity at $x_A$.
}\label{VelocityOne4x4}
\end{figure}

\textbf{Space, State and Time:} In EGT analogy~\cite{Sigmund2010},
 without loss of generality, we use a two-population MPG as an example. In the first population $X$, the strategy set is $\{X_{1}$,$X_{2}\}$ for each agent; similarly, in the second population $Y$, it is $\{Y_{1}$,$Y_{2}\}$. The payoff matrix is in shown Table~\ref{tab:MP5005}~\footnote{In each cell, the numeric in up-left is the payoff for the agent from $X$ and down-right for the agent from $Y$.}. If there are 4 agents in each population, an observable instantaneous populations strategy state should be  $x$:=$(p,q)\in \mathbb{X}$.  $\mathbb{X}$ is the populations strategy state space and  $\mathbb{X}$=$\{0,\frac{1}{4},\frac{2}{4},\frac{3}{4},1\}
 \bigotimes\{0,\frac{1}{4},\frac{2}{4},\frac{3}{4},1\}$, and $p$~($q$) is the density of $X_1$~($Y_1$) in $X$~($Y$). Figure~\ref{VelocityOne4x4} is an illustration, in which the square $5 \times 5$ lattice (gray dots) is the space  $\mathbb{X}$ and $x_{A}$=($\frac{1}{4},\frac{3}{4}$) is a state.

In experimental economics, the time $t$ is the label along the repeated rounds in a session. At each round $t$, there is one observation of $x_{t}$ in $\mathbb{X}$. Therefore, the \emph{smallest} tick $ \triangle t$ is $1$ and is the time interval between two successive rounds.

\begin{table} 
\caption{Payoff matrix of the MPG}
\centering
\begin{tabular}{|r|r|r|}
  \hline
    & $ Y_{1} $ & $ Y_{2} $ \\
  \hline
  $~ X_{1} ~$ & $~5~~~~~$ & $~0~~~~~$ \\
              & $~~~~~0~$ & $~~~~~5~$ \\
 \hline
  $~ X_{2} ~$ & $~0~~~~~$ & $~5~~~~~$ \\
              & $~~~~~5~$ & $~~~~~0~$ \\
  \hline
\end{tabular}
\label{tab:MP5005}
\end{table}

\textbf{Metric for Velocity:} Figure~\ref{VelocityOne4x4} shows a microscopic process $\mathbb{P}$ from which one observation of \emph{an instantaneous velocity}, $\upsilon^{o}_{pq}$, can be defined~\cite{XuWang2011ICCS}. From  $t$ to  $t$+$1$, the  forward change at a given state $(p,q)$ is described as $x^{+}$=$x_{t+1}-x_{t}$.
From  $t$-$1$ to  $t$, the  backward change is
  $x^{-}$=$x_{t}-x_{t-1}. $
Hence, at a given state $(p,q)$, we define, $\upsilon^{o}_{pq}$, as

\begin{equation}\label{eq:OneObsV}
 \upsilon^{o}_{pq} :=  \left(x^{+} + x^{-} \right)/ (2\triangle t).
\end{equation}
Practically, in experimental economics,  $\triangle t$=1.
In Eq.(\ref{eq:OneObsV}), $\upsilon^{o}_{pq}$ satisfies two
requirements for the measurement of instantaneous velocity (1) instantaneous: time interval is practically the smallest; (2) time odd: it is time reversal asymmetric~\footnote{Without lost of the generality, as shown in Fig.\ref{VelocityOne4x4}, if
$\mathbb{P}^T$ is the time reversal process of $\mathbb{P}$, the path of $\mathbb{P}^T$ remains the same as $\mathbb{P}$ but the direction of propagation is reversed~\cite{Meijer2004}, e.g., from $C$ to $A$ to $B$, and so, with Eq.~(\ref{eq:OneObsV}), the $v^{o}_{A}$ in $\mathbb{P}^T$ should be $(x_B - x_C)/2$  and is the opposite to $v^{o}_{A}$ in $\mathbb{P}$, so $\upsilon^{o}_{pq}$ is time reversal asymmetry.  
}. Then, the mean velocity vector at the $(p,q)$ can be computed as

\begin{equation}\label{eq:MeanObsV}
  \overline{\upsilon}_{pq} =  \frac{1}{\Omega_{pq}} \sum_{o} \upsilon^{o}_{pq},
\end{equation}
where $\Omega_{pq}$ is the total observation of the state $(p,q)$  and
the summation is carried over all of the rounds whenever the observed
state is  $(p,q)$. A practical numeric example for calculation  of a $\overline{\upsilon}_{pq}$ is shown in the \emph{S2} in electronic supplementary materials.

\textbf{Experiment:}
This experiment uses the traditional experimental setting for social evolution~\cite{Crawford1991,Binmore1993, Huyck1995,Binmore2001, HuyckSamuelson2001, cason2005dynamics,selten2008}.  We employed the MPG among two populations with the payoff matrix in  Table~\ref{tab:MP5005}. Each population includes 4 subjects. There are 12 independent sessions and each session includes 300 rounds repeated with random matching protocol.  For more details, see \emph{S1} in the electronic supplementary materials.

\begin{figure}
\centering
\includegraphics[angle=0,width=6cm]{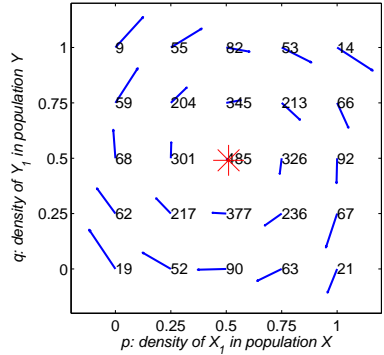}
\caption{Obtained evolutionary velocity field in laboratory MPG. The red star ($\ast$) indicates
 the mean observation of density. The numeric label indicates the observed occupation $\Omega_{pq}$ of each state.
The blue arrow from each state indicates the direction and the magnitude of mean velocity $\overline{\upsilon}_{pq}$. The plotting ratio is $1:1$.}
\label{MainMP}
\end{figure}

\textbf{Results:}
%
Figure~\ref{MainMP} shows the main results from our experimental data for the MPG.
The mean observation of density is $x^{*}$=$(0.510,0.490)$.
For each state, Figure 2 shows the total number of observation $\Omega_{pq}$  and the mean velocity vector $\overline{\upsilon}_{pq}$. Globally, a significant cyclic velocity field pattern is observed.
%

For state $(p,q)$, the \emph{normal vector} is defined as $r_{pq}$:=$x_{pq}-x^{*}$.  Our results show a statistically significant positive  correlation between  $|\overline{\upsilon}_{pq}|$ and $|r_{pq}|$ ($p$=0.000); The angle from $\overline{\upsilon}_{pq}$ to $r_{pq}$  is a right angle ($\pm 7\%$)
\footnote{Statistical analysis.
For the relationship of $|\upsilon_{pq}|$ and $|r_{pq}|$,  the statistical model of the simple linear regression is $|\upsilon_{pq}|=\beta~|r_{pq}| + \alpha$, where the samples are the 25 states cover all $(p,q)$ frequency weighted by $\Omega_{pq}$, and the OLS regression results are $\beta$=0.276$\pm$0.001 and $\alpha$=0.007$\pm$0.000. For the angle $\theta_{pq}$ from $\overline{\upsilon}_{pq}$ to $r_{pq}$ and $\theta_{pq}$:=$\arccos (\textbf{r}_{pq} \bullet \overline{\upsilon}_{pq})/(|\textbf{r}_{pq}| |\overline{\upsilon}_{pq}|)$, we have (1) using the 25 states cover all $(p,q)$ as the samples, in mean$\pm$s.e. format $\theta_{pq}$=1.614$\pm$0.062; (2) using the 25 states cover all $(p,q)$ as the samples but excluding the state (0.5,0.5), from $t$-$test$ with $Ho$:~mean=$\pi/2$, the results are mean value is 1.557, $Std.~Err.$=0.0282, $Std.~Dev.$=0.138 and the 95$\%$ Conf.~Interval is in [$1.499, 1.615$] and (3) from $ci$ test with weighted $\Omega_{pq}$  but excluding the state (0.5,0.5), the results are mean value is 1.584, $Std.~Err.$=0.0226 and the 95$\%$ Conf.~Interval is in [$1.538, 1.631$].
 We use software package Stata 10 to archive above results.}.

\textbf{Summary:}  In our data of MPG, statistically, the larger the magnitude of the normal vector, the faster the evolution should be. Meanwhile, the direction of the evolution is perpendicular to the normal vector.
In the full population strategy state space, the evolutionary velocity field pattern is cyclic and clockwise.

The existence of the velocity pattern is not observed only in our data. Using the same experimetrics as above and the data from ref.~\cite{selten2008} and ref.~\cite{Binmore2001},  which contains 19 different payoff matrix treatments of 2$\times$2 games of laboratory experimental economics, results of the significant cyclic velocity fields are als observed. For more details  see \emph{S3} in the electronic supplementary materials.

The matching pennies game is the simplest model in game theory~\cite{VonNeumann1944} and the evolution dynamical pattern can be modeled theoretically~\cite{Weibull1997,Sandholm2011}. In this letter,  we  report the first empirical observation of the evolutionary velocity field.

\textbf{Acknowledgment:} Thanks to B. O'Neill, K. Binmore, Y. Chen, R. Selten, S. Sunder, T. N. Cason, A. Roth, S. J. Goerg,  H. Wu, Y. Jia, Z. Dong, J. Wooders and Q. Cheng for the valuable comments and/or providing data.
Comments from ESA-PA 2011, ESA-EU 2011 conference and EEF Workshop in Xiamen University 2010
were helpful.

%

%

\bibliography{mainMar0517}
\end{document}